# Preparing Pre-College Students for the Second Quantum Revolution with Core Concepts in Quantum Information Science

Chandralekha Singh[1], Akash Levy[2], Jeremy Levy[1]

[1]University of Pittsburgh, Pittsburgh, PA

[2]Stanford University, Stanford, CA

After the passage of the US National Quantum Initiative Act in December 2018 [1], the National Science Foundation (NSF) and the Office of Science and Technology Policy (OSTP) recently assembled an interagency working group and conducted a workshop titled "Key Concepts for Future Quantum Information Science Learners" that focused on identifying core concepts for future curricular and educator activities [2-3] to help pre-college students engage with quantum information science (QIS). Helping pre-college students learn these key concepts in QIS is an effective approach to introducing them to the Second Quantum Revolution and inspiring them to become future contributors in the growing field of quantum information science and technology as leaders in areas related to quantum computing, communication and sensing. This paper is a call to pre-college educators to contemplate including QIS concepts into their existing courses at appropriate levels and get involved in the development of curricular materials suitable for their students. Also, research shows that compare and contrast activities can provide an effective approach to helping students learn [4]. Therefore, we illustrate a pedagogical approach that contrasts the classical and quantum concepts so that educators could adapt them for their students in their lesson plans to help them learn the differences between key concepts in quantum and classical contexts.

Introduction: While the crowning inventions of the first quantum revolution—transistors, lasers, computers—continue to enrich our lives, newfound excitement surrounds the use of quantum phenomena to create a new, second quantum revolution. The first quantum revolution was made possible by the formulation and development of quantum mechanics. The second quantum revolution is enabled by the exquisite coherent control and manipulation of tiny quantum systems that promise transformative improvements in our ability to compute, communicate and sense [5-17]. This new field of quantum information science and technology (QIST) promises new opportunities that led to the passing of the US National Quantum Initiative Act [1]. Shortly afterward, the NSF and OSTP assembled an interagency working group to focus on the "Key Concepts for Future Quantum Information Science Learners" particularly targeting pre-college education [2-3]. The key QIS concepts delineated for pre-college students produced by the interagency working group can be found in [2]. Although educators have focused on improving quantum education, most of these efforts have targeted college students (e.g., see Refs. [18-33]). However, to inspire future generations of QIST scientists and engineers, pre-college educators must play a key role in developing materials and engaging their students with core QIS concepts at appropriate levels in a variety of classes at different grade levels.

Contrasting classical and quantum concepts: Below we focus on a pedagogical approach that contrasts classical and quantum concepts relevant to QIS, which educators could adapt to suit their students' expertise [2-3]. One of us is working with the K-12 teachers to develop learning objectives and trajectories consistent with these key concepts [3]. We recognize that the adaptation of these concepts by the teachers of conceptual physics and AP physics into their lesson plans would look very different. The goal of this paper is not to help teachers develop a deep understanding of these concepts and if they are excited to pursue these topics further through AAPT quantum workshops promoted periodically in the eNNOUNCER since last year, this article has succeeded in its goals. We discuss a few of the most important concepts here to familiarize the teachers with these concepts, and leave discussion of others to the Appendix.

### State

A **classical state** of a physical system describes a particular condition that the system is in at a given time and is given by a collection of system properties. For example, "a ball at a certain location, moving in a certain direction with a certain speed", or even "heads" vs. "tails" of a coin. A successful physical theory should be able to describe states precisely, and predict how they change with time. For many purposes, we tend to gloss over irrelevant details about the state of a system. For example, the exact orientation of a coin lying flat on a surface is not relevant to the question of whether it is heads or tails. Let us consider a piece on a checker board. We can say that there are 32 different places on the checker board for the piece, and therefore 32 possible states for the system which consists of the checkerboard and the piece jointly. One can change the state by moving the piece from one place to another. Classical states change according to machinery that is encoded in the classical laws of physics, which can in turn be sculpted into the functioning of familiar machines that perform computations, including computers and smart phones. Computing machines possess a classical memory, along with the ability to change it in a programmatic way to achieve computation.

A **quantum state** is a mathematical description of the quantum system and contains all information that can be known about the system. If we return to the example of the checker board, we recall that there are 32 possible distinct states for the piece. The "quantum space" that describes all possible states for a single piece on a quantum checker board is 32 dimensional. What this means is that unlike classical checkers, a quantum checker piece can exist in all of the 32 possible locations at the same time. Being in multiple states simultaneously is often called being in a "quantum superposition" of states (more next). Scientists are now able to create such precise quantum systems which act like a quantum checkers game, and in fact now anyone can log into a cloud-accessible quantum computer [34] with 5 quantum bits (with $2^5$=32 distinct states) and put a quantum checker piece on all 32 sites at once!

### Superposition of states

Classically, there is no concept of being in a superposition of states. The closest idea to superposition might be the **superposition of classical waves**, which can interfere constructively or destructively. However, classical waves still have well-defined values.

The concept of quantum superposition is central to quantum mechanics in general and quantum information in particular. A quantum coin can be in a **quantum superposition** of both head and tail states at the same time. However, when measured (more later), it can only be found to be in two possible states, heads or tails with a certain probability. Similarly, a quantum top, in general, can be in a superposition of "spinning" simultaneously in clockwise and counterclockwise directions.

### Information (bits and qubits)

**Bits** are the smallest unit of (**classical) information**. All bits are stored in physical states of **matter** (or light) as one of two possible states or configurations. For example, a light switch can be in the "on" or "off" state. A coin can be heads or tails. We give binary assignments to the two states of a bit, "0" and "1", and assign them in a prescribed way to the physical states, e.g., "0"=heads, "1"=tails. Computers store information in many ways, often based on the amount of charge stored on a tiny capacitor. Information can also be stored as the North-South orientation of tiny magnets (in a hard disk drive), and in other well-defined states of matter.

**Qubits** or quantum bits are the smallest unit of **quantum information**. Qubits are embodied in forms of **quantum matter** (or light) that have two well-defined distinct states, e.g., the spin states of an electron or polarization states of a single photon. We label these distinct state "|0>" and "|1>". This **notation** has a precise mathematical meaning (they are like unit vectors), but the math is not required to explain the basic concepts to high school students. Unlike classical bits, qubits can exist in a quantum superposition. That means that if |0> and |1> are allowed states of the qubit, then any linear combination, such as |0>+|1> or |0>-|1>, is also a possible state.

## Measurement

**Measurement** is familiar in everyday life. You can measure whether a coin is heads or tails simply by looking at it. You expect that the coin was in a well-defined state just before you looked, and the act of looking at it has no effect on what you find.

**Quantum measurement** is a process whereby a given property of a quantum system is queried and becomes definite. The quantum system does not need to have a well-defined value before the measurement. A quantum measurement on a qubit "collapses" its quantum state onto one of two distinct outcomes, with relative likelihoods that depend on how the superposition is constructed. One important difference between measurement on classical bits and qubits is that you can measure more than one type of quantity with a qubit by asking different questions. Different quantum measurements are in this sense like asking different types of questions about a qubit. Some questions have definite answers, while others may be probabilistic. In particular, with classical bits, your measurement can only ask "are you zero, or are you one?" with the measurement outcome either being zero or one with 100% certainty. On the other hand, some of the questions you can ask about a qubit will yield definite answers, while others will give a probabilistic outcome. To be concrete, we can consider the spin of an electron, which has two distinct quantum states. However, you can only measure a component of the spin (which has three components similar to orbital angular momentum vector). Measurement of any particular component of the spin corresponds to a different question. Conventionally, the |0> state is assigned to the "spin-up" state, while the |1> state corresponds to the "spin-down" state. The question $Q_z$ asks, "is the qubit in the |0> state, or is it in the |1> state?". Thus, the measurement $Q_z$ inquires about the z-component of spin. However, we could instead ask the question $Q_x$, which asks "is the qubit in the |0>+|1> state, or is it in the |0>-|1> state?". If you are in the |0> state and ask question $Q_x$, then you will find that upon measurement, it is in the |0>+|1> state with 50% probability, and the |0>-|1> state with 50% probability. The same is true if we start with the |1> state. However, if you are in the |0>+|1> state when you ask $Q_x$, you will find the state after the measurement to be |0>+|1> every single time. Quantum measurement plays a central role in quantum theory and quantum information.

## Computing (information processing)

**Computers process information**. They do so by using **logic gates** which convert, e.g., two bits into one (e.g., via "AND" operation). By combining these logic operations, you can build up complex programs that run your phone, computer, etc. These logic gates in the computer can process bits very fast. However, some problems are too complex for the most powerful "classical" or "regular" computers on the planet. For example, while finding the factors of 15 is quick (3 times 5), finding the prime factors of a 400-digit number that is the product of two 200-digit prime numbers would take all the computers on the planet longer than the age of the universe to determine, using the best known classical algorithms. The world currently relies on this difficulty of factoring numbers to encrypt information over the internet.

**Quantum computers process quantum information.** They do so by using **quantum logic gates** that transform qubits or quantum states into other quantum states. For example, a commonly used single qubit quantum gate transforms $|0\rangle \rightarrow |0\rangle+|1\rangle$ and $|1\rangle \rightarrow |0\rangle-|1\rangle$. By combining this and other types of quantum logic gates, complex quantum algorithms can be implemented. A particularly important quantum algorithm (invented by the computer scientist Peter Shor [12]) can efficiently factor a 400-digit number in a reasonable amount of time. If a large quantum computer could be built today, it would be able to read all sensitive information circulating on the internet--credit card numbers, private messages, etc. Researchers are still far from this goal. There are many other constructive uses for quantum computers, such as helping to design new drugs, understand fundamental properties of our universe. The Second Quantum Revolution awaits the next generations of young students!

There are more concepts that can be taught to students than can fit within these pages. A few more key concepts, describing multiple bits/qubits, quantum entanglement, quantum key distribution, quantum teleportation, and internet/quantum internet are described in the Appendix.

Summary: We discuss some key QIS concepts [2-3] that pre-college educators could incorporate in their classes by adapting them as appropriate and illustrate these concepts using a pedagogical approach that contrasts classical/quantum cases. This paper is not trying to help pre-college educators develop a deep understanding of these concepts but it is a call to them to get involved in developing QIS curricular materials appropriate for their students. We have discussed and shared these compare and contrast approach with some K-12 educators and they have expressed interest in using this approach with their students. Some teachers have expressed that they would encourage their students to work in small groups and explore these key concepts further via searching for additional information on the internet and presenting to other students and their teacher in the class. We encourage K-12 educators to participate in AAPT quantum workshops promoted periodically in the eNNOUNCER.

# Appendix

## Multiple bits/qubits

All classical information can be broken down into sequences of bits, i.e., in general, the state of any classical complex system can be represented in terms of multiple bits. If we have 5 bits, then we can arrange each bit as 0 or 1 for a total of 32 distinct ways of arranging 5 bits. In general, every time you add a bit, you multiply the possibilities of distinct ways of arranging by two. Similarly, the number of distinct states in which a system of 5 qubits can be observed is the same as that for a classical set of 5 bits: 32.

Quantum information can also be stored in multiple qubits. However, by the principle of superposition in quantum case, it is possible for a system of 5 qubits to be in a superposition of all 32 states (or any number up to 32) at the same time. To represent a general superposition state in an ordinary computer, we would need to write down 32 numbers that ultimately would translate into the probability of each of the 32 outcomes of a measurement. If the number of qubits becomes too large (e.g., >300) then the number of numbers we would need to represent a quantum state in a classical computer would exceed the number of particles in the known universe. And yet physical qubits, e.g., spin states of electrons or polarization states of photons, keep track of this information for a living. That's why we need quantum matter to create large numbers of qubits. Later, we will see what even a modest number of qubits is capable of.

## Entanglement

**Classical Entanglement** comes from the word "tangle" which can be illustrated by the following everyday scenario: suppose you and your friend store your ID cards in the same bowl. If you accidentally pick up your friend's ID, your friend is likely to pick up yours. These kinds of classical correlations happen when the opportunity to interact arises.

**Quantum entanglement** is novel and has no known classical analogue. In entangled states, the outcome of the measurement of one qubit is correlated with that of another qubit but before the measurement you cannot predict which qubit will yield what outcome. If we have two qubits, then there are four distinct states which we can label as |00>, |01>,|10>, |11>. The meaning of these new symbols, e.g.,|10>, is that the first qubit is in the |1> state and the second qubit is in the |0> state. There are superposition states such as |00>+|01>+|10>+|11> that look complicated but in fact are not entangled in the sense that there is a 50% chance of measuring |0> or |1> for either qubit, and there is no correlation between what is measured for the first qubit and what is measured for the second qubit. However, there are other superposition states, for example, |01>+|10> or |01>-|10>, that are entangled because they are correlated in special ways. Either of these two states has the property that if the first qubit is found to be |0>, the second would be |1>, and vice-versa. However, we cannot predict the outcome of a particular measurement ahead of time, and we can only claim that it will either be |01> or |10>. What is truly bizarre is that we can physically separate the two entangled qubits over large distances, and this quantum correlation would still be true. The implications of this "quantum entanglement" were deeply disturbing to Einstein, but they help form the basis for a new "quantum internet" that has yet to be built but will offer new security capabilities not possible with our current internet. Entanglement will also play a key role in building a scalable quantum computer.

## Key distribution

Encryption "keys" are central to sending information securely over the internet. **Key distribution** is employed for encrypting and decrypting information. A key is a randomly chosen string of bits (e.g., K=01101011) that both sender and receiver share. To encrypt a "message" (e.g., "11110000"), you could decide, e.g., to align the key with the message and toggle the bit of your message only if the corresponding key bit was "1". The encrypted message would be 10011011. Then this message would be transmitted over a public channel, e.g., the internet to the receiver but anyone else could also intercept it. However, if only the receiver has the same key, then only they can convert the message back to the original information by performing the same operation to decrypt the message. We share, e.g., our credit card information over the internet using encryption schemes. But how do you share the encryption key securely with the receiver so that only they can decrypt the encrypted message you send over a public channel? Classically, generating a shared key for secure transmission of information with someone with whom you have not previously met is not possible. Protocols exist that enable two parties to distribute a key that is secure, as long as it is difficult to factor large numbers into their primes.

One way to avoid the potential security breach from a quantum computer is to use another quantum invention: **Quantum key distribution** (QKD). There are several protocols, but the most popular approaches these days make use of non-orthogonal polarization states of photons. These photons are shared between two people who wish to securely communicate with one another in such a way that the photons cannot be intercepted by an eavesdropper without their presence being detected. The first protocol was developed by Bennett et al. in 1984 [6], and a subsequent version was developed in 1992 [7-11]. QKD protocols are being adopted by banking industries for secure communication between different banks that are nearby [14]. We are close to achieving large-scale quantum communication, but are limited by the distance that we can send qubits without destroying the quantum information due to decoherence and losses. More work is needed to create and preserve these entangled states over large distances.

## Teleportation

**Teleportation** literally means transmission, but sadly we are not referring to what routinely happens on Star Trek. Every time you look something up on the internet, you are teleporting classical information from a storage location "in the cloud" to your browser. We are not transmitting cups or people but rather just information or descriptions.

**Quantum teleportation** refers to sending the quantum state of a qubit (e.g., $|0>+|1>$ or $|0>$) from one place to another. The difficulty is that if you try to convert the quantum information to classical form, you have to perform a measurement by asking a precise question and you destroy the quantum nature of the state. For example, if you ask question $Q_z$ for state $|0>$, then you will always get $|0>$ and you can reliably transmit that over a classical internet. But if you ask question $Q_z$ for state $|0>+|1>$, then half the time you will get $|0>$ and half the time you get $|1>$. So you will be sending different information randomly even though it is just one definite state $|0>+|1>$, that you intended to send. You could, instead, ask question $Q_x$ for the state $|0>+|1>$, and that would always give you the answer $|0>+|1>$. But if you asked question $Q_x$ for state $|0>$, then again you would half the time get $|0>+|1>$ and half the time $|0>-|1>$. Measurement destroys the quantum information you intended to send. Thus, you have to somehow send the quantum bit without measuring it.   There are ways to do this and they involve entanglement, which can be thought of as a "resource".  If you share a pair of entangled qubits with a friend, then you can use this pair of qubits as a resource to send a qubit that you have to your friend.  This is called "quantum teleportation".

## Internet

The **internet** is a set of links between many "nodes" (computers or servers) along with rules for governing how information is transmitted, "securely" and redundantly. A quantum internet does not exist yet, but it would be a way to send quantum information between many nodes, along with rules for governing how that quantum information is transmitted, securely and redundantly.

A **quantum internet** would presumably be built up of a network of nodes that are sharing quantum information via quantum teleportation.  A quantum internet would solve the security challenges we have with our existing internet, and could also enable distributed quantum computing and many other wondrous things.